\documentclass[10pt, conference, letterpaper]{IEEEtran}
\usepackage{tikz,amsmath,amssymb}
\usetikzlibrary{shapes,backgrounds}
\usepackage{verbatim}
\usepackage{comment}
\usepackage{algorithm}
\usepackage[font=small,labelfont=bf]{caption}
\usepackage[font=scriptsize]{subcaption}
\usepackage[noend]{algpseudocode}
\usepackage{multirow}
\usepackage{pifont}
\usepackage{balance}
\usepackage{enumitem}
\usepackage{color, colortbl}

\usepackage{pifont}
\newcommand{\cmark}{\ding{51}}%

\definecolor{Gray}{gray}{0.9}
\newcolumntype{g}{>{\columncolor{Gray}}c}
\begin{document}
    
    \title{On Network Traffic Forecasting using Autoregressive Models}
    
    \author{\IEEEauthorblockN{Do\u{g}analp Ergen\c{c}}
    
        \IEEEauthorblockA{University of Hamburg, Germany \\ ergenc@informatik.uni-hamburg.de}
   		\\
			\IEEEauthorblockN{Ertan Onur}
			 \IEEEauthorblockA{Middle East Technical University, Turkey \\ eonur@metu.edu.tr}}

    \IEEEtitleabstractindextext{%
    \begin{abstract}
Various statistical analysis methods are studied for years to extract accurate trends of network traffic and predict the future load mainly to allocate required resources. Besides, many stochastic modeling techniques are offered to represent fundamental characteristics of different types of network traffic. In this study, we analyze autoregressive traffic forecasting techniques considering their popularity and wide-use in the domain. In comparison to similar works, we present important traffic characteristics and discussions from the literature to create a self-consistent guidance along with the survey. Then, we approach to techniques in the literature revealing which network characteristics they can capture offering a characteristic-based framework. Most importantly, we aim to fill the gap between the statistical analysis of those methods and their relevance with networking by dicussing significant aspects and requirements for accurate forecasting from a network-telemetric perspective.
  
    \end{abstract}
    \begin{IEEEkeywords}
     time-series, forecasting, stochastic modeling, autoregressive
    \end{IEEEkeywords}}
        
    \maketitle
    
    \IEEEdisplaynontitleabstractindextext
    
    \IEEEpeerreviewmaketitle

\section{Introduction}

In the early years of the Internet, network traffic had been modeled with relatively easy statistical approaches. There were only a few commonly-used services and protocols, and they are actively used by a very limited number of users. In contrast, it is much harder to predict traffic patterns and characteristics in today's communication systems. Even if many different techniques are embodied for analysis and prediction, several concerns  must be addressed for an accurate traffic engineering. Time-series models are quite popular to extract the temporal patterns of network traffics and make predictions depending on those patterns \cite{hyndman2014forecasting}. 

There are different approaches for forecasting such as exponential smoothing \cite{Roman2013, Tikunov2007}, wavelets \cite{Wang2002, Zang2015} and hybrid methods including multiple approches \cite{Ouyang2017, Yongtao2011}. Besides, neural networks (NNs) and autoregressive models are two of frequently-used group of techniques for network traffic prediction in practice. They are considered as the fundamental elements of forecasting toolbox. Today, network traffic forecasting with NN is quite popular as a different approach than traditional stochastic modeling \cite{Yu1993, Feng2015, Zhuo2017}. They detect patterns and structures in input data, learn through many iterations and use such experience to evaluate new data similar to learning process of human beings. NNs are more successful to capture complex relationships in data thanks to their non-linear nature but the data required for training is much higher in comparison to autoregressive models. Even if NNs seem like the primary alternative for them, autoregressive methods are dominantly studied especially for the prediciton of network traffic excluding other domains.

\begin{figure*}[h!]
        \centering
        \includegraphics[scale=.45]{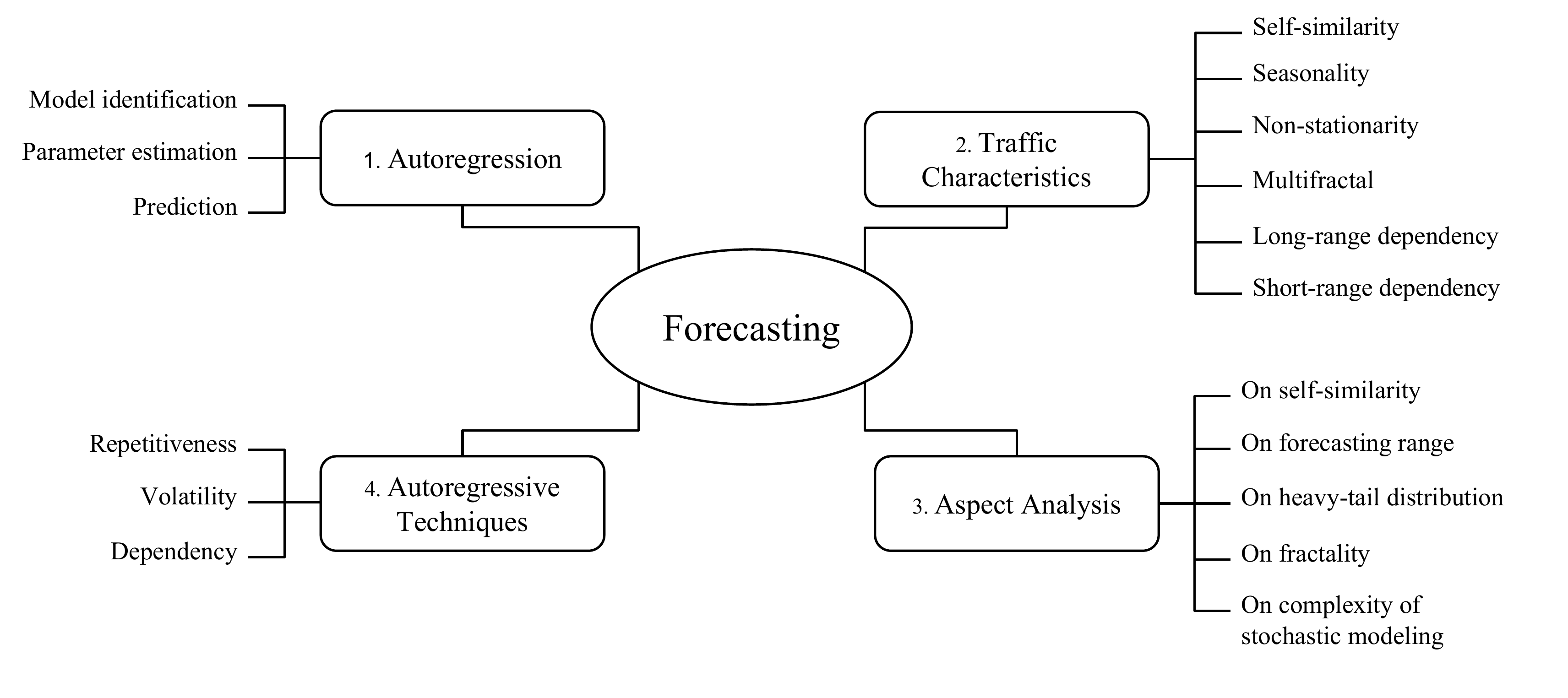}%
        \caption{A summarizing picture for the whole study.}
        \label{fig:mindmap}
    \end{figure*}

To the best of our knowledge, there are only a few surveys in the literature on network traffic forecasting \cite{Joshi2015, Bolshinsky2012} and the existing ones do not even touch the significant network flow aspects. Besides, there is not a systematical study that builds a grounding for traffic forecasting research offering an analysis framework. In this study, we present a self-consistent study that analyzes requirements, characteristics, and examples of temporal autoregressive models for forecasting since they are mostly employed and practically used models for network traffic prediction. Rather than examining the statistical foundation of the models, we review all aspects of forecasting from a higher-level networking perspective. Fig. \ref{fig:mindmap} shows a mindmap that summarizes all important headlines of the study. Accordingly, our contributions are listed as:

\begin{itemize}
\item We review the relevant dynamics of autoregressive modeling techniques which are common in various studies (Section \ref{sec:tech}).
\item We discuss different characteristics of time-series data from networking perspective for a better comprehension of the forecasting studies rather than touching analytical details (Section \ref{sec:char}). Moreover, we use such characteristics as a framework to analyze forecasting studies.
\item We present a short analysis of different aspects of traffic flows (Section \ref{sec:analysis}) and analyzed various autoregressive studies concerning about such aspects and characteristics (Section \ref{sec:practical}). We also group the studies under a more general meta-framework apart from the characteristics.
\item We point out common issues and challanges, and also possible research directions in general (Section \ref{sec:discussion}).
\end{itemize}

\section{Brief Introduction to Autoregressive Modeling} \label{sec:tech}

\begin{figure}[h!]
        \centering
        \includegraphics[scale=.40]{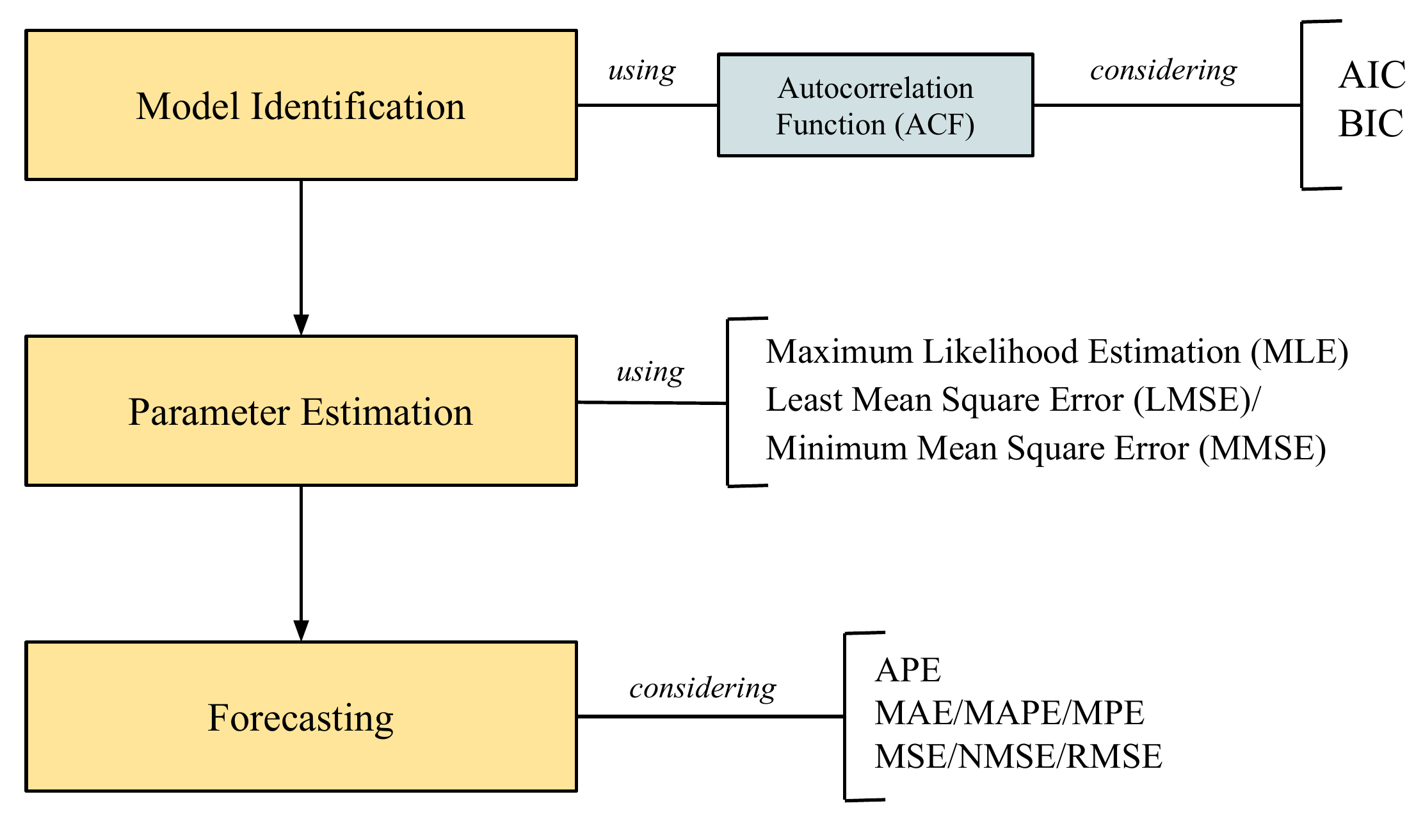}%
        \caption{Abstract block diagram for the dynamics autoregressive models.}
        \label{fig:arframework}
    \end{figure}

Autoregressive (AR) models are stochastic models that consume the input values (of past) in a time sequence into a regression function to predict future values for related time-series. Autoregressive Moving Average (ARMA) \cite{Cadzow1983}, Autoregressive Integrated Moving Average (ARIMA) \cite{Box2015}, Fractional ARIMA (FARIMA) \cite{Yang1999}, Seasonal ARIMA (SARIMA), Autoregressive Conditional Heteroskedasticity (ARCH) \cite{Chen2006}, Generalised ARCH (GARCH) \cite{Engle1995}, Exponential GARCH (EGARCH), Autoregressive Conditional Duration (ACD) \cite{Engle1998}, Stochastic Autoregressive Mean (SAM), and Nonlinear Auto Regressive with Exogenous (NARX) are falling into this category. Indeed there are statistical differences, for instace, while ARIMA models focus on conditional mean through temporal series, ARCH methods take conditional variance into consideration for modeling. In this study, we especially focus on those techniques under autoregression scope considering their practicality, relatively shorter modeling duration, less data requirement, and lower complexity. The dynamics of those models is simple shown in Fig. \ref{fig:arframework} for a better understanding. Note that, it is a very brief illustration and omits iterative processes which are required for optimization of the model. 

AR models consist of three phases: (i) Statistical modeling with respect to some criteria, (ii) parameter estimation and (iii) forecasting. (i) The first phase is related to detect correlation in time-series using autocorrelation functions (ACFs) and identify the model based-on widely-used criteria such as Akaike Information Criterion (AIC) and Bayesian Information Criterion (BIC) \cite{Ding2018}. (ii) Secondly, related coefficients of the identified model are estimated using well-known estimation methods such as Maximum Likelihood Estimation (MLE) and Least Mean Square Error (LMSE). After parameter estimation, (iii) future-points of the time-series are predicted and the accuracy is presented with respect to different metrics. 

Lastly, there are a number of metrics to evaluate the performance of time-series modeling techniques. All of the presented studies in Section \ref{sec:practical} employs one or more metric(s) to analyze the accuracy of their proposed techniques and compare with other techniques. Generally, after the model is constructed, such metrics are used to compare fitted (i.e., predicted values according to the model) and actual values. Table \ref{tab:metric} shows those metrics briefly and they are associated to the studies presented in next sections. 

\begin{table}[htbp]
\caption{List performance metrics commonly used in different studies. If a metric is uniteless, it is defined as a ratio or percentage independent by its actual measuring unit. Scale-free, on the other hand, represents normalized metrics.}
\label{tab:metric}
\centering
\resizebox{\columnwidth}{!} {
\begin{tabular}{c|c|c|c}

\textbf{Abbreviation} & \textbf{Metric} & \textbf{Uniteless} & \textbf{Scale-free} \\ \hline
APE & Absoulute percentage error & \cmark & \cmark \\ \hline
MAE & Mean absolute error &  & \\ \hline
MARE & Mean absolute relative error & \cmark  & \cmark \\ \hline
MAPE & Mean absolute percentage error & \cmark & \cmark \\ \hline
MPE & Mean percentage error & \cmark & \cmark \\ \hline
MSE & Mean square error & & \\ \hline
NMSE & Normalize mean square error & & \cmark \\ \hline
NRMSE & Normalize root mean square error & & \cmark \\ \hline
RMSE & Root mean square error &  &\\ \hline
SER & Signal-to-error ratio & \cmark & \cmark\\ \hline
\end{tabular}
}
\end{table}

Since dozens of studies analyze the statistical aspects of AR models in depth in various domains, we are not taking this approach here and keep related discussion limited with this section. The terms touched here may be considered as a guidance for a better comprehension of the rest of this study, especially in Section \ref{sec:related}.

\section{Traffic Characteristics} \label{sec:char}

Network traffic may reflect various characteristics that are vital to detect for accurate forecasting. Those characteristics are not only observed in network traffics but any subject analyzing temporal data such as economics, physics, and psychology. In this section, we introduce six characteristics from a network-telemetric perspective. Those characteristics also construct our comparison framework to analyze and compare network traffic forecasting studies in Section \ref{sec:practical}.

\textbf{Self-similarity.} Being introduced 40 years ago \cite{Mandelbrot1982}, self-similarity of network traffic is discussed in a number of studies \cite{Erramilli2002, Park2000}. In a very practical sense, self-similar objects are observed in the same shape when magnified or shrunk. From the network traffic perspective, it means that a proportional segment of measuring (e.g., number of packets or amount of data during a certain of time with a predefined granularity) tends to be observed in a different time scale. In this sense, self-similarity is also related to long-range dependency in practice. 

\textbf{Seasonality.} It is quite common to observe nearly the same patterns with a certain frequency in any domain of temporal measuring. Typical weather conditions in each "season" of a year are great examples. Similarly, for network traffics, weekend and weekdays, holidays or certain hours of a day show very similar patterns, periodically. Comprehension of the seasonality in data flows is crucial to analyze the nature of traffic, and also forecast the future traffic load possibly reflecting congruent patterns.

\textbf{Non-stationarity.} In a stationary process, the mean, variance, and correlation model stay constant over time and it is one of the very common assumptions for time-series modeling and stochastic processes. However, network traffic may show changing statistical characteristic leading to a change in modeling as well \cite{Karagiannis20042, Cao2001, Morris2000}. Therefore, before forecasting, a time-series model should be capable and sensitive to detect such changes that are observed frequently depending on various factors such as the number of users, connections, and bandwidth utilization of related network elements in practice.

\textbf{Multifractal.} In aggregated network traffics (i.e., consisting of multiple flows originated by multiple sources), it is possible to observe self-similar characteristics of individual network flows. This type of traffic flows are not only indicated as self-similar or fractal but multifractal \cite{Taqqu1997}. Detecting multifractal behaviors and fractal patterns of multiple flows simultaneously are naturally important yet challenging for forecasting.

\textbf{Long-range dependency (LRD).} Various time-dependent systems or physical phenomena show correlated behavior during large time scales. For instance, Hurst confirmed such a situation in Nile River's repeating rain and drought conditions observed a long period of time \cite{Hurst1956} and Hurst parameter becomes the fundamental detection technique of the LRD. For network traffic, especially the Internet, the long-range dependency is observed in different significant studies \cite{Grossglauser1999, Karagiannis2004}. Together with the self-similar properties, the LRD shifts the traffic modeling perspective from memoryless stochastic processes (e.g., Poisson) to long-memory time-series.

\textbf{Short-range dependency (SRD).} In comparison to the LRD, the correlation can be observed for shorter time scales in short-range dependent processes. That is, the dependence among the observations quickly dissolves and it is related to quickly decaying correlations. Many traditional time-series modeling techniques examine the SRD; however, it is not enough to reflect today's network traffic pattern. SRD can be considered while forecasting short-term traffic employing relatively low-complexity models.

Note that those characteristics are mutually complementary. For instance, LRD and self-similarity require similar methods to be detected but they are not the same: the former is strongly related to time-dependency while the latter one covers the variance in scale as well. However, the common case is being dependent on scaling patterns in long time intervals. It is also possible to analyze such characteristics through multifractal traffic analysis. Similarly, the seasonality actually infers the non-stationary characteristics but while some studies directly consider seasonality, the others are looking for the non-stationarities from a wider perspective. Therefore, we present those elements separately as they are individually addressed in various works in the literature.

\section{Forecasting in the Literature} \label{sec:related}

After a high-level introduction on the autoregressive models and the definition of traffic characteristics, we relate all that background to forecasting methods in the literature discussing a number of significant aspects. In this section, first, we present further key aspects that fundamentally reshape network traffic analysis pointing to significant studies. Then, we review various studies that exemplify what is covered so far.

\subsection{Analysis on Different Aspects of Traffic} \label{sec:analysis}

Before presenting the temporal techniques for traffic modeling, we introduce some short discussions about the different aspects of network traffic. We selected those aspects since they are considered as milestones in forecasting domain that change research perspective and lead to more accurate forecasting or comprehension to underlying reasons for the limited success in flow prediction. Through this section, we evaluate important studies on the analysis of such aspects for a better understanding and further thinking on forecasting.

\textbf{On self-similarity.} Self-similarity was a complex issue and significantly required novel stochastic techniques to be considered in forecasting. In \cite{Leland1993}, the nature of self-similarity in Ethernet traffic is discussed both practically and statistically. The authors analyzed self-similarity relying on rescaled range analysis (through Hurst parameter), variances in aggregated processes and periodogram-based analysis in frequency domain. It is revealed that the stochastic processes to model Ethernet traffics are not capable to reflect their self-similar nature which is proven with statistical analysis on a four-year captured Ethernet traffic. Moreover, the authors suggest Hurst parameter is much effective to detect burstiness in traffic in comparison to known parameters such as the index of dispersion, peak-to-mean ratio, and coefficients of variation. Lastly, they list two ways to generate network traffic satisfying self-similarity property: fractural Gaussian noise (FGN) and chaos maps. The study is quite revealing (though its age, +25 years) to deeply understand the underlying statistical properties of the real Ethernet traffic and also the analysis of self-similarity.

\textbf{On forecasting range.} It is important to estimate the limits of success in forecasting and define requirements precisely for accurate resource allocation for the potential traffic. In \cite{Sang2002}, the authors focus on three fundamental issues of the traffic forecasting: (i) how far into the future a traffic can be predicted, (ii) how much resources are required to minimize uncertainty and (iii) which characteristics of traffic are the most effective. For (i), they define the maximum prediction interval (MPI) with having a limited prediction error $e$ (e.g., 20\%) with a certain probability $p$ (e.g., 99.9\%) during an interval of time. Besides, the tradeoff between maximizing MPI and minimizing prediction error is well-discussed. (ii) In most of the referenced studies here, first- and second-order statistics of the traffic are considered and there needs to be a certain amount of resource for accurate statistical modeling. (iii) According to the study, the prediction efficiency may not be directly related to the selected method (which are ARMA and MMPP here) but the nature of traffic. For instance, while the ethernet traffic is not predictable with a certain error for a reasonable MPI, the study shows much better performance on Internet traffic due to the higher multiplexing factor. That is, it is not efficient to take sample traces (e.g., sessions) for prediction if the traffic is fed from many different (i.e., statistically independent) sources exactly like the Internet. Similarly, the efficiency in the use of certain characteristics needs to be discussed per scenario. The study draws the conclusion that the LRD may not matter in traffic management for delay-sensitive services, accordingly.

\textbf{On heavy-tail distribution.} Heavy-tail distribution in network traffic is shown to be strongly related to the transfer size and interarrival times with the self-similar nature of the (Internet) traffic \cite{Crovella1997}. Heavy-tail distribution is defined as having a heavier tail than the exponential distribution \cite{Bryson1974}. It easily misleads the traffic forecasting methods relying on basic statistics. In \cite{Fapojuwo2006}, the main focus is on detection and characterization of heavy-tail distribution. The existing estimators are not accurate to reflect heavy-tail characteristics since they have idealized assumptions such as stationarity and independence. Therefore, first, the authors present quantile-quantile and complementary cumulative distribution function (CCDF) plot for the detection of heavy-tails indicating the drawbacks of those methods. Then, the performance of four different estimators for the characterization (i.e., detecting the tail exponent) and their sensitivity to noise are analyzed. To compensate for their weaknesses, the authors propose a new wavelet-based method to filter long-range dependent data and increase the efficiency of previously used estimators. In the end, the homogeneity of long-range variance (i.e., time-varying LRD exponent) is discussed. Even if the study does not directly present a forecasting method, it is worth pointing out to understand the effects of heavy-tail distribution and LRD for the analysis of time-series data. 

\textbf{On fractality.} Self-similar traffic patterns are widely detected using Hurst parameter; however, it is rather complicated to analyze multifractal flows following the same procedure. Complex (and high-speed as stated in the study) networks generally comprise multiple self-similar and independent flows that can be modeled as different stochastic processes and cumulative analysis of such fractal flows (i.e., considering as a single flow) shows a multifractal behavior. In \cite{Millan2016}, the authors relate multifractal patterns of the flow to the locality phenomenon of Hurst parameter in self-similar networks. It is also indicated that varying Hurst parameter would be the key method to generate traffic flows with multiple self-similar characteristics. 

\textbf{On complexity of stochastic modeling.} Without inspecting for correlation in any scale, it is questioned that if we can model various network traffics using simpler approaches e.g., Poisson distribution in packet interarrival times. It indeed directly depends on the nature of the traffic. For instance, according to \cite{Paxson1995}, modeling TCP traffic with Poisson (or other models) cannot capture LRD and burstiness and eventually results in degrading performance of forecasting in terms of average packet delay or maximum queue size. That is, a deeper analysis and more sophisticated modeling are required to represent such traffic e.g., TCP in wide-area networking. On the other hand, high aggregation on the Internet (for instance a traffic sample captured from the backbone traffic) may nearly follow Poisson distribution \cite{Terdik2009} and it eventually indicated the weakness of bursty traffic in the backbone traffic. 

\subsection{Autoregressive Modeling Techniques} \label{sec:practical}

There are a number of studies that uses time-series statistical methods by modifying them according the nature of the applied network traffic. In this section, we present how they modify or enhance the techniques in Section \ref{sec:tech} to satisfy various requirements. Table \ref{tab:prediction} shows comparative analysis taking different network characteristics into consideration. One can argue that the basic dynamics of some methods address several characteristics by default. For instance, while differencing in ARIMA is a solution to eliminate non-stationarity, ARMA automatically detects short-range patterns. However, a characteristic is marked for a study in Table \ref{tab:prediction} only if related study directly addresses a problem related to that particular characteristic specifically and its effectiveness is shown using required measurement techniques.

    \begin{figure*}[h!]
        \centering
        \includegraphics[scale=.40]{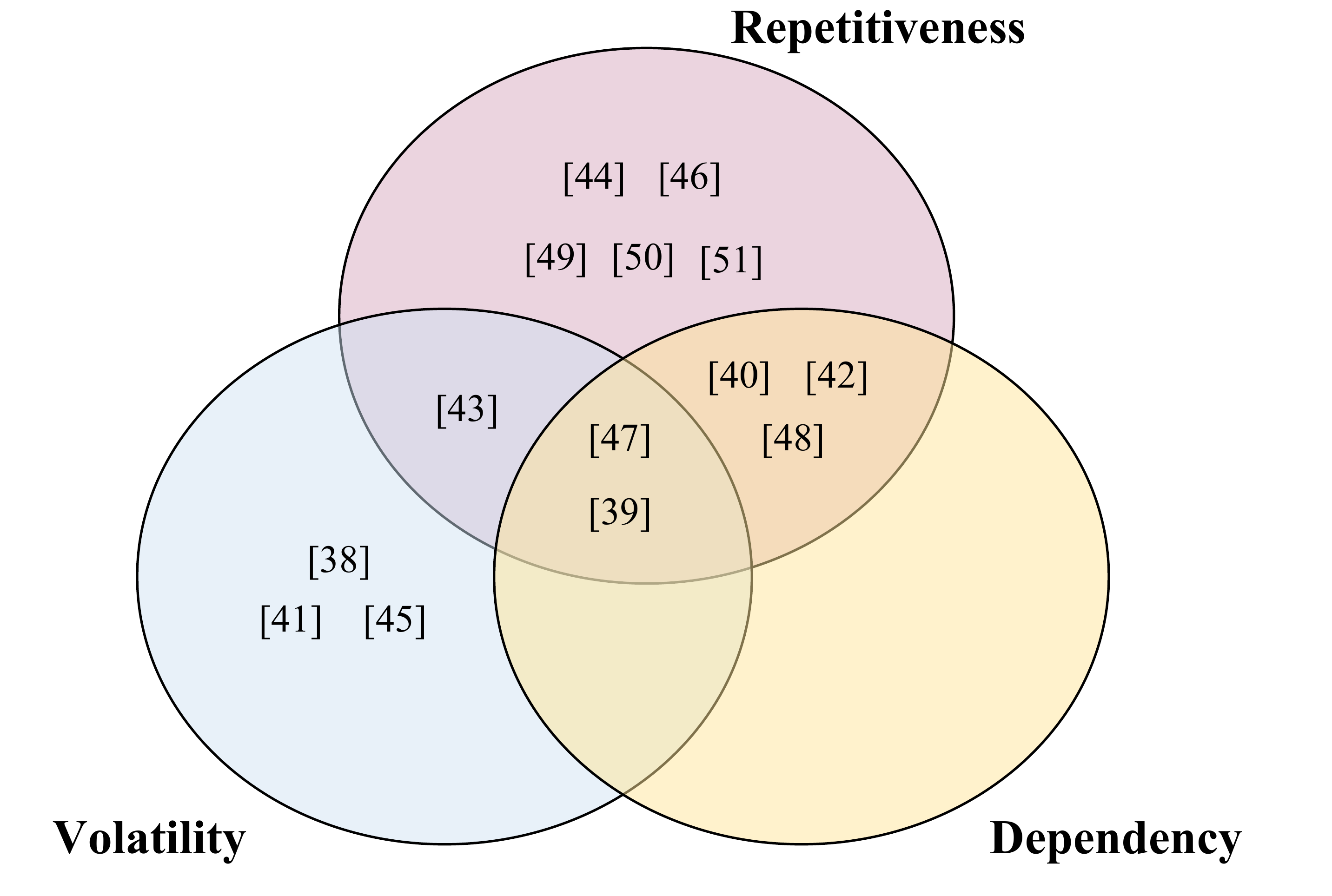}%
        \caption{The studies are divided into three main focal groups, dealing with repetitiveness, volatility, and dependency of network traffic.}
        \label{fig:artypes}
    \end{figure*}

The studies here take similar statistical approaches but focus on different characteristics. It is possible to divide them into three groups in terms of such characteristics to handle, repetitiveness, volatility, and dependency. \textit{Repetitiveness} represents cyclic and usual patterns and it is directly related to self-similarity and seasonality. \textit{Volatility}, on the other hand, covers the varying properties of network traffic such as non-stationaries and multifractals. Lastly, \textit{dependency} represents the time-dependent characteristics which are long- and short-range dependency. This classification can be considered as a meta-framework corresponding to the characteristics presented in Section \ref{sec:char}. Fig. \ref{fig:artypes} groups the studies in this section according to this meta-framework.

As seen in Fig. \ref{fig:artypes}, many studies have commonly addressed multiple characteristics. It is also reasonable to explain cross-relationships between categoricals through the figure. For instance, \textit{repetitiveness} and \textit{dependency} may be very closely-related since they both lead to temporal correlations. However, while \textit{dependency} mainly focuses on timing issues, \textit{repetitiveness} is also related to scaling in magnitude with an observable pattern. \textit{Volatility} may require different analysis based-on decomposition but still has intersections between the others since it targets instability in both time and scale. Addressing the problems on all three categories is possible, though. Such studies are generally hybrid methods and expected to have higher complexity. In the rest of this section, the studies falling into those categories are discussed. Note that since half of them belongs to multiple groups, they are not gathered under individual headlines and sorted in chronological order.

\begin{table*}[htbp]
\caption{List of autoregressive modelling and prediction methods.}
\label{tab:prediction}
\centering
\renewcommand{\arraystretch}{2} 
\resizebox{\textwidth}{!}{
\begin{tabular}{c|cccc|cccccc}

\textbf{Study} & \textbf{Technique(s)} & \shortstack{\textbf{Parameter}\\ \textbf{Estimation}} & \textbf{Evaluation} & \textbf{Domain} & \shortstack{\textbf{Self-}\\ \textbf{similarity}} & \textbf{Seasonal}  &  \shortstack{\textbf{Non-}\\ \textbf{stationarity}} & \textbf{Multifractal} & \textbf{LRD} & \textbf{SRD}  \\ \hline

Mao \cite{Mao2005} & \shortstack{ARIMA\\ Wavelet} & \shortstack{MLE\\ LS}  & \shortstack{MARE\\ NMSE} & LAN & &  & \cmark  & & &\\ \hline
Zhou \textit{et al.}\cite{Zhou2006} & \shortstack{ARIMA\\ GARCH} &  \shortstack{ACF\\ MLE}  & SER & WAN & \cmark  &  & \cmark  & & \cmark  &\\ \hline
Vujicic \textit{et al.}\cite{Vujicic2006}&  SARIMA & - & NMSE & \shortstack{Public\\ Safety} &   & \cmark  &   & &   &\\ \hline
Jun \textit{et al.}\cite{Jun2007} &  \shortstack{ARMA\\MLSL}  & LSL & MSE & Internet & \cmark &   &  & & \cmark &  \cmark \\ \hline
\shortstack{El Hag and\\ Sharif}\cite{ElHag2007} &  AARIMA & - & MAE & WAN & \cmark &    &  & & \cmark &   \\ \hline
Anand \textit{et al.}\cite{Anand2008} & GARCH & MLE & NMSE(-like)  & Internet &  &  & \cmark  & &&  \\ \hline
Chen \textit{et al.}\cite{Chen2009} & SARIMA &  \shortstack{ACF\\ LS} & MAPE & WLAN & & \cmark & &   && \cmark \\ \hline
Yu \textit{et al.}\cite{Yu2011}  & ARIMA & MLE & MAPE & \shortstack{Cellular\\ Network} &  &\cmark  &\cmark & &&   \\ \hline
Yu \textit{et al.}\cite{Yu20112} & APM & MLE  & MAPE & \shortstack{Cellular\\ Network}  && \cmark  & &&&  \\ \hline
\shortstack{Zhang and\\ Huang}\cite{Zhang2013} &  \shortstack{ACD\\ Particle Filter}  & BHHH  & RMSE & Data Center &  & & \cmark && & \\ \hline
Hu \textit{et al.}\cite{Hu2013} &  \shortstack{X-12 ARIMA\\STL}  & -  & APE  & SNMP && \cmark & && &   \\ \hline
Yu \textit{et al.}\cite{Yu2013} & FARIMA &  \shortstack{MLE\\MMSE} &  \shortstack{APE\\ MAPE} & \shortstack{Cellular\\ Network}  & \cmark &  &   &\cmark & \cmark  & \\ \hline
Yimu \textit{et al.}\cite{Yimu2015} & \shortstack{GARCH\\ LMD}  &  MLE &  RMSE & \shortstack{P2P\\ Multimedia}  & \cmark &  &  &&\cmark  & \\ \hline
\shortstack{Yoo and\\ Sim}\cite{Yoo2016} & \shortstack{ARIMA\\STL}  & MLE  & \shortstack{RMSE\\ MAE}   & SNMP & \cmark & \cmark &   & & &  \\ \hline
Markovic \textit{et al.}\cite{Markovic2017} & SARIMA  & Manual &  \shortstack{RMSE\\ MAE} & Multimedia  &&\cmark  &&  &&  \\ \hline
Xu \textit{et al.}\cite{Xu2018} & NARX-RF & L-BFGS  & APE  & \shortstack{Cellular\\ Network} & &\cmark & && & \\ \hline

\end{tabular}
}
\end{table*}

It is quite likely that different network mechanisms cause traffic variations at different timescales. Therefore, it is generally hard to statistically model network traffic at once. In \cite{Mao2005}, the authors decompose the traffic into different timescales using \textit{a-trous} Haar wavelet transform. Then, they apply different ARIMA models to each wavelet (i.e., traffic at different timescales) for one-step forecasting at seconds granularity. Combining the prediction for each, the forecast traffic which covers varying characteristics of the traffic through time is obtained. The results are compared with an NN-based method using the very same training data and it is shown that the proposed method shows better performance in terms of NMSE and MARE. 

ARIMA/GARCH \cite{Zhou2006} offers a combined technique targetting various network characteristics such as SRD, LRD, self-similarity and multifractal. Using the abilities of ARIMA in linear traffic and GARCH for changing variance, the authors design a one-step predictor which shows a potential to be extended to make multi-k (or multi-step) predictions. Similar to \cite{Anand2008}, ARIMA/GARCH has a parameter estimation phase to tune the parameters of both ARIMA and GARCH using MLE based on the Box-Cox \cite{Box1964} method. The technique shows better performance than FARIMA in terms of signal-to-error ratio (SER) in various timescales and also experimental multi-k predictions. However, the authors do not present concrete accuracy results other than SER. Therefore, the correlation between predictions and actual traffic data is not directly observable.

In \cite{Vujicic2006}, instead of analyzing aggregated traffic, the authors profile user behaviors based on their hourly call-rate using K-means clustering method. They divide users (call-groups in the study context) into three groups as low, medium and high call-rate by the clustering method. Then, each group is separately modeled using SARIMA models for daily and weekly cyclic patterns and then add them up to make complete prediction reflecting all users' behavior. The main motivation is taking advantage of well-defined group characteristics for more accurate predictions rather than the whole data which is relatively harder to model due to its complex and aggregated nature. However, it is not feasible to work on maximum granularity (i.e., per-user prediction) though. Therefore, the authors aim to provide (a degree of) scalability by grouping the users while increasing the prediction accuracy in comparison to modeling aggregated traffic. The results show that 57\% of group-based predictions gives better results than the predictions on aggregated traffic in terms of NMSE. Besides, group-based forecasting paves the way of profiling individual users as far as they can be identified under a forecasting group.

\cite{Jun2007} proposes a modified version of Least-square lattice (MLSL)\cite{Alexander1986} method to calculate related autoregression (AR) parameters dynamically. MLSL relies on adaptive filter theory. Instead of evaluation of model parameters once using a set of training data, it dynamically updates AR model parameters per input (i.e., packet). The authors modified LSL to reduce computation cost and increase convergence speed. In comparison to least-squared method (LS) and ARMA, MLSL shows higher accuracy and faster convergence experimenting on synthetic data that have short-term dependence characteristics. For data generation, the authors use inverse Fourier transform to generate Fractal Gaussian Noise given Hurst parameter. Therefore, the data show self-similarity as well (as SRD).

\cite{ElHag2007} proposes the Adjusted ARIMA Model (AARIMA) for modelling Internet traffic data at millisecond time scales. The authors speficially address self-similarity and LRD in Internet traffic whose samples are captured from Bellcore Internet Wide Area Network. They have shown that even if the residuals of ARIMA models give residuals with a white noise distribution, the models may not offer sufficient goodness of fit statistics. Therefore, they offer AARIMA as a quick and simple modeling method by modifying ARIMA where the first difference of the stationary series added as a regressor. Especially for modeling Internet traffic, the results show that AARIMA gives lower MAE finalizing in higher number of iterations in different datasets. Note that in terms of modeling phases presented in Section \ref{fig:arframework}, evaluating AARIMA is exactly the same with ARIMA as specifically claimed by the authors and it satisfies all requirements for reliable residuals (e.g., white noise distribution, Box-Jenkins tests).

In \cite{Anand2008}, an enhanced (or generalized) ARCH model (GARCH) is introduced to develop a one-step predictor for non-linear traffic models e.g., Internet. The authors point out that constant-variance models like ARIMA and its successors (e.g., FARIMA and SARIMA) cannot fit the bursty (and non-linear) nature of the Internet traffic whereas GARCH is taking conditional variance into account to react changing traffic patterns. To be able to determine GARCH parameters, MLE is deployed using the training data. The results show that the forecast error of GARCH is significantly less than the ARIMA-ARCH model for one-step prediction (i.e., comparing to ARIMA(1,1,1)-ARCH(1)). However, its performance is open to validation in less aggregated traffics other than the Internet.

Despite its simplicity, \cite{Yu2011} offers a leaner ARIMA differencing process by converting multiple stationarization operation (for both trend and seasonal patterns) to a single multiplicative process. The authors speficially target to eliminate seasonal patterns (i.e., analyzing 12-months (or lag) autocorrelation results) and also extracting 6-month patterns from a large network traffic data based-on a set of Chinese regions. They present single- and multi-step prediction MAPEs for each month; however, the results are not compared any other prediction method that is applied to related data.  A similar approach is also presented in \cite{Chen2009} namely Multiple Seasonal ARIMA model to obtain a model covering two different seasonal features in wireless network traffic using 5-minutes sampling.

In consideration of relatively complex and time consuming process of ARIMA, accumulation predicting model (APM) is proposed in \cite{Yu20112}. APM especially addresses the traffic patterns with stable seasonal characteristic. For the detection of stable seasonality, the ratio of partial accumulation to total accumulation (i.e., cumulative traffic load for a certain amount of months to the whole year) is evaluated and constant (e.g., linear in the study) changes are interpreted as a reflection of the characteristic. Then, such interpretation is used to predict monthly traffic for the next year. When it is compared to ARIMA, APM results in lower MAPE in the detection of the total monthly traffic; however, ARIMA is more successful to estimate average daily traffic (with relatively higher APE deviation).

It is usually quite hard to model and forecast network traffic in minute-granularity. In \cite{Zhang2013}, the authors propose an autoregressive model for short-term prediction, in minutes. Autoregressive conditional duration (ACD) \cite{Tsay2009} is employed to model the distribution of interarrival times of continuous traffic for a certain duration. Related parameters of ACD is estimated using Berndt-Hall-Hall-Hausman (BHHH) algorithm. The traffic, as claimed by the authors, is non-stationary and has a non-Gaussian distribution. Then, the particle filtering method is applied considering the extracted distribution model for the prediction of the packet traffic in the upcoming minute. The results show that the proposed method is able to predict one-minute traffic (in MBs) with less error than ARMA in terms of RMSE.

Detecting seasonality is one of the crucial tasks for decompositon of the traffic. In \cite{Hu2013}, the adjustment of seasonality is detailedly investigated to understand underlying clearly and make more accurate forecasting. For the adjustment, the authors handle missing values first and soften outliers dividing them into four different types. Then, using seasonal-trend decomposition using Loess (STL) \cite{Cleveland1990} and X12-ARIMA \cite{Findley1998}, they decompose the data into three components: trend, season and irregularities. An extra step, diagnostics, is taken to examine the stability of adjusted data series with a persistent model for new data feeds. It is revealed that considering a daily seasonality leads to most accurate decomposition and show the trend with minimum irregularities after this procedure. The results show that the one-step forceasting after cleaning dailiy seasonality on Simple Network Management Protocol (SNMP) data gives the minimum forecasting error in comparsion to six other benchmarking methods including Holt-Winters \cite{Holt2004, Winters1960}, ARIMA and linear regression. Another similar study working on SNMP data is conducted addressing the seasonality \cite{Yoo2016}. After seasonal adjustment using STL, the authors also analyze self-similarity. Lastly, they apply ARIMA to remove residual autocorrelation on the adjusted data. The results show the success of the proposed study on forecasting network utilization under stationary and non-stationary assumptions, and varying size of training set with one-day seasonal cycles.

Even if short- and long-range dependency in various types of network traffic are discussed, fractal characteristics may be harder to detect. FARIMA, for instance, is designed to handle fractal patterns considering inadequency of ARIMA. In \cite{Yu2013}, the authors discuss the points where FARIMA fails, the multifractal characteristics. When hourly traffic of sequential days is examined in mobile 3G downlink traffic, it is shown that self-similarity and multifractal patterns exist and FARIMA would not be enough to model such patterns for the prediction. Therefore, a combined technique of FARIMA and ARIMA is embodied to eliminate (a) fractal, (b) long-range dependent and (c) short-range dependent characteristics successively. Finally, an effective method examining the change in Hurst parameter for predicted data is used for forecasting. The results show that the combined method results in less than 8\% APE and nearly 2\% MAPE that are considered as reasonable error rates for daily forecasting.

One of the significant outcomes of traffic forecasting is increasing the quality of service (QoS) by predicting user demand. In \cite{Yimu2015}, the authors analyze peer-to-peer (P2P) video sharing to satisfy QoS requirements in multimedia traffic. They, first, determine the fundamental problems of such traffic which are claimed as self-similarity and LRD in the study. To deal with those characteristics, the whole temporal data of network traffic is divided smaller-term time series, or Product Functions (PFs), using local mean decomposition (LMD) \cite{Smith2005}. LMD helps to process the series as multiple short-range series independent by its long-term patterns. Then, several PFs are iteratively forecast using GARCH and the predicted values are summed up to obtain a final prediction. The results are compared ARMA and WNN models in terms of RMSE and it is shown that proposed method offer more accurate predictions. 

After high-definition videos have changed multimedia trends, upcoming 4K/8K videos are expected to become popular in the same direction. However, increasing resolution in videos naturally affects required network resources to watch them online. In \cite{Markovic2017}, the authors analyze the nature of 4K video traffic for modeling and forecasting so that resource allocation can be handled in advance. They experiment with multiple SARIMA models with various modeling parameters to extract both seasonal and non-seasonal patterns. The best model is selected with respect to AIC and optimized minimizing RMSE and MAE without using a parameter estimation method. Further optimization and model fitting are examined using Ljung-Box tests \cite{Box2015} and Empirical Cumulative Distribution Function (ECDF) graphs. The results show that 4K videos have frequently changing frame size variance without a certain pattern and therefore it is not possible to make long-term predictions. 

The authors in \cite{Xu2018} address a very practical problem: The varying network use in holidays especially for cellular networks. It is highly possible that people are communicating in special days and holidays much more frequently than ordinary days and it directly affects the quality of service of such periods. In cellular networks, the estimation of changes in traffic patterns in those days per base station is important for service providers to arrange required resources. In \cite{Xu2018}, eliminating long and seasonal trends of traffic, the traffic patterns in holidays are extracted for each base station and they are clustered using K-means clustering. Then, the data in each cluster are modeled using random forest (RF) method to obtain the relationship between the input variables and the traffic data of similar patterns (in the same cluster). This model is also used for the prediction, combining with Nonlinear Auto Regressive with Exogenous model (NARX-RF) whose parameters are estimated using Limited-memory Broyden-Fletcher-Goldfarb-Shanno (L-BFGS) algorithm \cite{Liu1989}. Therefore, the study takes adventage of time-series, unsupervised and supervised learning methods all together. It is compared with Facebook's Prophet \cite{Taylor2017} and the results show that it has significantly lower absolute percentage error (APE).

\section{Discussion and Conclusion} \label{sec:discussion}

In this study, we propose a guideline and make a broad discussion about traffic forecasting with autoregressive methods focusing on the network-related issues rather than the statistical analysis. We especially pay attention to make the whole study consistent touching to various characteristics and grounding different studies in the literature on a framework depending on such characteristics. In the last section, we are highlighting a wrap-up discussion to cover the most general issues of forecasting that we extract from all studies presented here.

It is clearly understood that \textit{offering a general purpose model is not possible} even if the network characteristics are quite similar. More than 50\% of the studies we reviewed majorly address seasonality in network traffic. Even if the most obvious seasonal patterns such as holidays and festivals are well-captured, some other cyclic patterns still need to be modeled more carefully. In such cases, the size of training data and optimization and estimation steps of autoregressive algorithms presented in Section \ref{sec:tech} have become crucial. That is, a modeling problem actually consists of multiple optimization problems whose results reshape the intended model and this issue leads to uncountably different models with varying parameters. 

In certain cases, even a single well-identified model becomes insufficient to make accurate predictions and the solution consequently \textit{converges to combined or hybrid models}. It increases complexity and required time for training, and decreases flexibility of the models as they rely on various dependent parameters. This issue also leads to development or employment of different techniques on network traffic analysis such as machine learning and neural networks. Indeed \textit{emerging techniques} bring their own issues on the table and it strongly becomes a trade-off between complexity and accuracy. Moreover, when other performance metrics such as prediction range and confidence interval are included, the coverage of the optimization problem gets beyond the limits. Related to that, it is considerable to \textit{involve heuristics and field expertize to narrow analytical problems to more practical ones}. For instance, Facebook's Prophet \cite{Taylor2017} offers forecasting ecosystem rather than a prediction technique called "analysts-in-the-loop" where the experts can directly involve the forecasting process instead of a fully-automatized prediction. Therefore, aside from sufficient statistical knowledge, it is also valuable to understand domain-specific network requirements to success.

Lastly, the common requirement for all forecasting methods is sufficient training data and proportional training time. Therefore, there is a \textit{huge necessity for both practical and real-time techniques that can be dynamically trained and reshaped} using spontaneous data. It may not be possible for the complex nature of the network traffic but research on sufficient heuristics that either ease the training process or increase the performance of forecasting alongside the statistical modeling. Autoregressive models, in this sense, are relatively easier to comprehend and be evolved. 

In summary, we aim to fill the gap between the statistical analysis of autoregressive forecasting methods and their relevance with networking by discussing significant aspects and requirements for accurate forecasting from a network-telemetric perspective. Even if we focus on the autoregressive methods in the survey part, we believe that our discussion of network traffic forecasting is conducted in a much more broader sense. For future work, we intend to expand the survey with more modern methods such as machine learning and neural networks.

\balance
    \bibliographystyle{IEEEtran}
    \bibliography{IEEEabrv,references}
    
\end{document}